\documentclass[prl,twocolumn,groupedaddress,amsmath,floatfix]{revtex4}
\usepackage{bm}
\newcommand{\phdag}{{\phantom{\dagger}}}

\newcommand{\kf}{k_{f}}
\newcommand{\vf}{v_{f}}

\newcommand{\be}{\begin{equation}}
\newcommand{\y}{g_{bs}}
\newcommand{\ybare}{\tilde{g}_{bs}}
\newcommand{\g}{g_{\rm fr}}
\newcommand{\gbare}{\tilde{g}_{\rm fr}}

\newcommand{\ee}{\end{equation}}
\newcommand{\bea}{\begin{eqnarray}}
\newcommand{\eea}{\end{eqnarray}}
\newcommand{\bse}{\begin{subequations}}
\newcommand{\ese}{\end{subequations}}

\begin{document} 
\title{Quantum decoupling transition in a one-dimensional
  Feshbach-resonant superfluid}

\author{Daniel E.~Sheehy and Leo Radzihovsky} 
\affiliation{Department of Physics,
  University of Colorado, Boulder, CO, 80309} 

\date{May 26, 2005}
\begin{abstract}
  We study a one-dimensional gas of fermionic atoms interacting via an
  s-wave molecular Feshbach resonance.  At low energies the system is
  characterized by two Josephson-coupled Luttinger liquids,
  corresponding to paired atomic and molecular superfluids. We show
  that, in contrast to higher dimensions, the system exhibits a
  quantum phase transition from a phase in which the two superfluids
  are locked together to one in which, at low energies, quantum
  fluctuations suppress the Feshbach resonance (Josephson) coupling,
  effectively decoupling the molecular and atomic superfluids.
  Experimental signatures of this quantum transition include the
  appearance of an out-of-phase gapless mode (in addition to the
  standard gapless in-phase mode) in the spectrum of the decoupled
  superfluid phase and a discontinuous change in the molecular
  momentum distribution function.

\end{abstract}
\maketitle

Recent experimental advances~\cite{expts} have led to a realization of
paired superfluidity in degenerate atomic gases. It is driven by the
atomic Feshbach resonance (FR) through a molecular state whose rest
energy (detuning) $\nu$ can be adjusted with a magnetic field. The
associated high tunability of interactions allows one to explore
superfluidity in these systems ranging from the BCS regime of strongly
overlapping Cooper pairs (for large positive detuning) to the BEC
regime of dilute Bose-condensed molecules (for negative
detuning)~\cite{theory}.

\begin{figure}[bth]
\vspace{2.2cm}\hspace{.08cm}
\centering
\setlength{\unitlength}{1mm}
\begin{picture}(40,40)(0,0)
\put(-52,0){\begin{picture}(0,0)(0,0)
\includegraphics{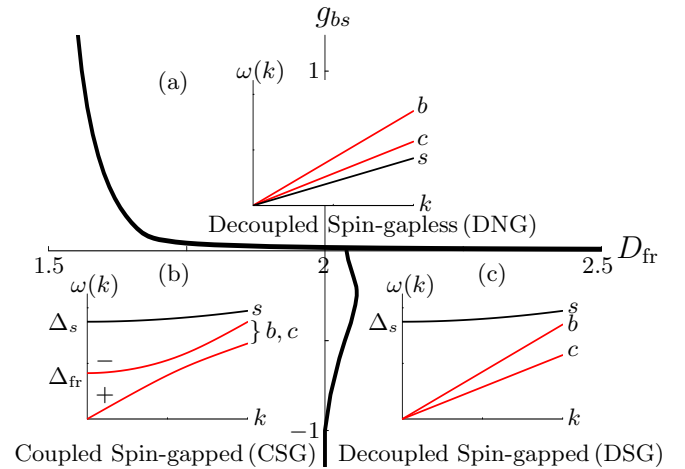}
\end{picture}}
\put(56,30.5) {\large${D_{\rm fr}}$}
\put(16,62) {\large$\y$}
\put(15.0,54.5) {$1$}
\put(12.5,6.5) {$-1$}
\put(-21.5,28.5) {$1.5$}
\put(16.25,28.5) {$2$} 
\put(51.5,28.5) {$2.5$}
\put(1.75,34) {Decoupled Spin-gapless \!(DNG)}
\put(-5,53) {(a)}
\put(29.5,49.75) {$b$}
\put(29.5,45.5) {$c$}
\put(29.5,43) {$s$}
\put(29.75,36.75) {$k$}
\put(5.5,54.5) {$\omega(k)$}
\put(-24,3.5) {Coupled Spin-gapped \!(CSG)}
\put(-5.0,27.5) {(b)}
\put(-13.25,11.5) {$+$}
\put(-13.25,16.0) {$-$}
\put(7.25,19.85) {$\}\,b,c$}
\put(7.5,23) {$s$}
\put(7.75,8.25) {$k$}
\put(-16.5,26) {$\omega(k)$}
\put(-19.5,21) {$\Delta_s$}
\put(-19.5,13.75) {$\Delta_{\rm fr}$}
\put(19,3.5) {Decoupled Spin-gapped \!(DSG)}
\put(37.5,27.5) {(c)}
\put(49.5,23.25) {$s$}
\put(49.5,20.75) {$b$}
\put(49.5,17.5) {$c$}
\put(49.5,8.25) {$k$}
\put(25.5,26) {$\omega(k)$}
\put(23,21.0) {$\Delta_s$}
\end{picture}
\vspace{-.3cm}
\caption{(Color online) Main: Phase diagram as a function of the scaling
  dimension $D_{\rm fr}$ of the FR coupling and the direct atomic
  interaction strength $\y$ for $\g = 0.19$.  For attractive
  interactions, $\y<0$, fermion spins are gapped and the transition is
  between coupled and decoupled spin-gapped paired superfluids.
  For repulsive atomic interactions, $\y>0$, the transition is between
  the coupled spin-gapped state and a decoupled spin-gapless state.
  Insets a,b,c: Plots of spectrum for fermion spin ($s$), fermion
  charge ($c$), and boson charge ($b$) excitations within these three
  phases.}
\label{fig:phasediagram}
\vspace{-.3cm}
\end{figure}

In all BEC-BCS crossover studies to date, it has been tacitly
(correctly~\cite{comment,Hulet}) assumed that the superfluid phases of
the closed-channel Bose-condensed molecules and open channel
Cooper-pairs are locked together by the FR coupling, with the
superfluid at low-energies characterized by a single gapless
Bogoliubov (0th sound) mode.  In this Letter we show that in striking
contrast, a one-dimensional (1d) resonantly interacting atomic
gas~\cite{Recati,RecatiPRA,Cazalilla03}, realized by a sufficiently
high aspect ratio trap~\cite{Paredes,Esslinger}, can exhibit a more
interesting possibility. Namely, for a range of parameters quantum
fluctuations, enhanced by the low dimensionality, suppress the FR
coupling thereby leading to a quantum phase transition into a
superfluid state where the Cooper-pair and molecular superfluids are
decoupled. Our main findings are summarized by the phase diagram in
Fig.~\ref{fig:phasediagram}. One striking experimental signature of
this decoupling transition is the appearance of an out-of-phase
gapless mode (in addition to the abovementioned standard gapless
in-phase mode) in the spectrum of the decoupled superfluid phase, that
should be observable through Bragg spectroscopy~\cite{bragg}.
Another signature is a jump across the transition in the exponent
$\alpha$ characterizing the molecular momentum distribution function
$n_b(k)\propto k^{-\alpha}$ with $\alpha_{\rm decoupled\/} <
\alpha_{\rm coupled\/}$, measurable via time-of-flight
images~\cite{Paredes}.

The decoupling transition can be seen through the bosonization
representation of the molecular and atomic quantum fluids as acoustic
charge and spin collective modes, with the FR interaction reducing to
a Josephson coupling between phases of the molecular and Cooper-pair
superfluids.  Now, a sufficiently strong atomic repulsion (attainable
through optical lattice
engineering\cite{Paredes,Esslinger,OpticalLattices}), that \lq\lq
localizes\rq\rq\ atom number can be tuned to enhance the
canonically-conjugate superfluid phase fluctuations, to the point that
the FR (Josephson) coupling is averaged away at low energies.

This transition has strong connections to other interesting examples
of fluctuation-driven decoupling of Josephson-coupled XY-models, that
fall into the roughenning universality class\cite{Jose}.  Most
notably, these include the sliding phases of DNA-lipid
complexes\cite{Ohern} and their quantum higher-dimensional Luttinger
liquid (LL) generalizations\cite{Emery00}, the latter a realization of
a long-sought-after LL power-law phenomenology in $d>1$.

We now demonstrate and explore the superfluid decoupling transition
through a sketch of our detailed calculations\cite{detailedPRB}. The
appropriate model of a two-species $\sigma=\uparrow,\downarrow$ (two
distinct values of an internal degree of freedom, e.g., hyperfine
states) fermionic atomic gas interacting through a molecular FR is
described by a Hamiltonian\cite{theory} $H = \sum_{k,\sigma}
\epsilon_k c_{k\sigma}^{\dagger} c_{k\sigma}^{\phdag}+ \sum_q
\big(\epsilon_q/2 + \nu)b_q^\dagger b_q^\phdag + H_{\rm fr} + H_{\rm
  int}$, where $\epsilon_k = \hbar^2 k^2/2m$, $\nu$ is the FR
detuning, and $H_{\rm int}$ is the direct fermionic and bosonic
interactions.

The FR interaction
\be H_{\rm fr} = -\gbare\int d^d x \big(c^{\dagger}_\uparrow(x)
c^{\dagger}_\downarrow(x) b(x) + h.c.\big),
\label{fbr}
\ee 
describes the interconversion between a pair of open-channel fermionic
atoms $c_{k\sigma}$ and a closed-channel diatomic molecule $b_q$.  The
form of this atom-molecule coupling guarantees that the Cooper-pair
and molecular superfluids must condense together, and that, in $d>1$,
tuning through an s-wave FR leads to a smooth BEC-BCS crossover
(rather than a phase transition) between two limits of a single paired
superfluid state.

However, in 1d long-range order (BEC) is precluded even at $T=0$ by
enhanced quantum fluctuations~\cite{Hohenberg}, and superfluidity is
characterized by a quasi-long-ranged superfluid order, that admits a
richer set of possibilities.  To see this, it is convenient to use a
bosonized representation~\cite{Haldane,Giamarchi} of the atomic and
molecular fields, with
\bea
b(x)&\sim& {\rm e}^{i\phi_b(x)},
\label{eq:bs}
\\
c_\sigma(x)&\sim&{\rm e}^{i \kf x} {\rm
e}^{i(\phi_\sigma(x)+\theta_\sigma(x))} + {\rm e}^{-i \kf x}{\rm
e}^{i(\phi_\sigma(x)-\theta_\sigma(x))},
\hspace{0.5cm}
\label{eq:cs}
\eea
an approximate form sufficient here. In the above $\kf$ is the Fermi
wavevector, the two terms in Eq.~(\ref{eq:cs}) are the right and left
moving contributions to $c_\sigma$, and $\theta_\sigma$ and
$\phi_\sigma$ are the fermion charge-density wave (phonon) and
superfluid phase fields, respectively.  These can be conveniently
split into charge $(c)$ and spin $(s)$ components via $\phi_{c,s} =
(\phi_\uparrow \pm \phi_\downarrow)/\sqrt{2}$ and similarly for
$\theta_{c,s}$.

Using Eqs.(\ref{eq:bs},\ref{eq:cs}) inside $H$ and focussing on
densities incommensurate with the optical lattice, we find the
coherent-state imaginary-time action (with $\hbar = 1$)
\bea 
&&S =  \int dx d\tau \Big(\sum_{\alpha = b,c}
\frac{K_\alpha}{2\pi} \big[ v_\alpha (\partial_x \phi_\alpha)^2 +
v_\alpha^{-1}
(\partial_\tau \phi_\alpha)^2 \big]\nonumber \\
&&\quad +\frac{1}{2\pi K_s} \big[ v_s(\partial_x \theta_s)^2 +
v_s^{-1} (\partial_\tau \theta_s)^2 \big] + \frac{\y}{2\pi a^2} \cos
\sqrt{8} \theta_s \nonumber
\\
&&\quad - \frac{\g}{\pi a^2} \cos(\phi_b\!-\!\sqrt{2} \phi_c)
\cos \sqrt{2}\theta_s \Big),
\label{eq:wholeaction}
\eea
governing the dynamics of three acoustic modes ($\phi_b$, $\phi_c$,
and $\theta_s$) coupled by two key nonlinearities $\y$ and $\g$, that
arise from the backscattering part of fermionic short-range
interaction $\ybare n_\uparrow(x) n_\downarrow(x)$ and from the FR
coupling, respectively. Here, the dimensionless couplings $\y =
\ybare/\pi v_f$ and $\g = 2\gbare \sqrt{\rho_b}/\pi \vf \rho_f$, with
$\rho_f$ and $\rho_b$ the detuning-dependent fermion and boson
densities, $\vf = \pi \rho_f/2m$ the Fermi velocity.  All other
relevant atomic and molecular interactions have been incorporated into
the Luttinger parameters $K_b$, $K_c$ and $K_s$, and velocities $v_b$,
$v_c$ and $v_s$. For weak interactions $K_{s,c}\approx 1\pm \y/2$ and
$K_b\to \infty$. In the above $a=(\pi \vf^{1/2} \rho_f)^{-1}$ is the
ultra-violet (UV) cutoff.

Based on experience with the sine-Gordon model~\cite{Jose,Giamarchi}
we expect and find that over a large range of Luttinger parameters
$K_\alpha$ a pertubative treatment of the backscattering and FR
nonlinearities breaks down.  A renormalization group (RG) treatment is
therefore necessary to ascertain the low-energy behavior of the
system. This amounts to successively integrating out the high-energy
degrees of freedom in an infinitesimal wavevector shell
$a^{-1}e^{-\ell} < q < a^{-1}$ around the UV cutoff. Resulting
effective couplings appearing in $S$ then determine the low-energy
thermodynamics of the system.

Before discussing the full behavior it is convenient to consider a
limiting regime of the above model. The simplest case is that of large
attractive atomic interactions, $\y\ll -1$ (that, as we will see
below, is induced by a finite FR scattering even if the nonresonant
atomic interaction $\y$ is moderately repulsive). In this limit, the
backscattering nonlinearity \lq\lq freezes\rq\rq\ $\theta_s$ at $0$,
corresponding to a spin-gap state with gap
$\Delta_{bs}\approx 2\sqrt{|g_{bs}|v_s K_s}/a $.
Using $\theta_s=0$ inside $S$ reduces the problem to two acoustic
modes $\phi_{b,c}$ coupled by a single FR nonlinearity. Its effect can
be assessed using an RG analysis that here amounts to a determination
of the scaling dimension $D$ of the FR operator ${\cal O}(x)\equiv
\cos(\phi_b(x) - \sqrt{2}\phi_c(x)) \cos \sqrt{2} \theta_s(x)$ around
the Gaussian fixed point, $\y=0, \g=0$.  This in turn is determined by
the long-scale behavior of $\langle {\cal O}(x) {\cal O}^{\dagger}(0)
\rangle \sim |x|^{-2D}$.  A simple calculation
in the spin-gap phase gives $D=D_{\rm fr}\equiv(4K_b)^{-1} +
(2K_c)^{-1}$. A standard RG analysis~\cite{Jose} then shows that for
$D_{\rm fr} < 2$ ($D_{\rm fr} > 2$) the effective FR coupling grows
(diminishes) at low energies under the RG coarse-graining procedure.
For $D_{\rm fr} < 2$ and weak $g_{\rm fr}\ll 1$, on scales longer than
$\xi_{\rm fr}\approx a v_f^{1/2} \g^{-\frac{1}{2-D_{\rm fr}}}$, the FR
coupling dominates over the kinetic energy and the growth of $\g$
saturates at $\g^{\frac{2}{2-D_{\rm fr}}}$. In this strong-coupling
regime, the FR interaction locks the closed-channel molecular
superfluid phase $\phi_b$ to the open-channel Cooper-pair superfluid
phase $\sqrt{2}\phi_c$, a characteristic of the \lq\lq{\em coupled}
spin-gap\rq\rq\ (CSG) state.
Approximating the FR coupling by a harmonic spring
$(\g^{\frac{2}{2-D_{\rm fr}}}/2\pi a^2) (\phi_b - \sqrt{2}\phi_c)^2$,
with a \lq\lq stiffness\rq\rq\ softened by quantum fluctuations up to
the scale $\xi_{\rm fr}$, and diagonalizing the quadratic form in
$\phi_b$ and $\phi_c$, gives two dispersions of the CSG state:
\hspace{-3cm}\bea &&\omega_\pm(k)^2 = \frac{1}{2}
\Big[(v_c^2+v_b^2)k^2 +
\Delta_{\rm fr}^2 \nonumber \\
&&\quad \mp\sqrt{[(v_c^2+v_b^2)k^2 + \Delta_{\rm fr}^2]^2 -
  4v_c^2v_b^2k^4 - 4c^2 k^2 \Delta_{\rm fr}^2 } \Big],\quad\quad
\label{eq:spectrum}
\eea
with $\Delta_{\rm fr}\equiv a^{-1}|\g|^{\frac{1}{2-D_{\rm fr}}}
\sqrt{v_b K_b^{-1}\!+\!2 v_c K_c^{-1}}$, and $c \equiv \sqrt{( v_c K_c
  + 2v_b K_b )/(v_c^{-1}K_c + 2v_b^{-1} K_b)}$.  The dispersion
$\omega_+(k)$ ($\approx c k$) characterizes the gapless Bogoliubov
mode corresponding to in-phase oscillations of the closed-channel
molecular ($\phi_b$) and the open-channel Cooper-pair ($\phi_c$)
superfluid phases.  The dispersion $\omega_-$ ($\approx \Delta_{\rm
  fr}$) is for the gapped mode (in the long wavelength limit given by
$\phi_-\equiv\phi_b - \sqrt{2}\phi_c$) in which $\phi_b$ and $\phi_c$
fluctuate out phase.  Hence, the CSG state is characterized by two
gapped modes $\theta_s, \phi_-$ and a single in-phase gapless
Bogoliubov mode (see Fig.~\ref{fig:phasediagram}b). In this coupled
state the molecular phase $\phi_b$ is characterized by an effective
Luttinger parameter $\bar{K}_b= \sqrt{(v_b K_b + v_c K_c/2)(v_b^{-1}
  K_b + v_c^{-1} K_c/2)} > K_b$ that can be read off from the action,
Eq.~(\ref{eq:wholeaction}) after simply imposing the low-energy FR
coupling constraint $\phi_c = \phi_b/\sqrt{2}$ inside $S$. Similarly,
atomic charge correlations are controlled by $\bar{K}_c=2\bar{K}_b$.

Now consider $D_{\rm fr}>2$ but still deep within the spin-gapped
state, $\y\ll-1$.  Here, quantum fluctuations of $\phi_{b,c}$ become
sufficiently strong so as to average away long-scale effects of the FR
coupling, reducing it relative to the kinetic energy of fluctuations
on these long scales. We refer to this distinct thermodynamic state
(special to 1d) as the \lq\lq{\em decoupled} spin-gap\rq\rq\ (DSG)
superfluid. It is characterized by effectively independent
fluctuations of the low energy molecular and Cooper-pair superfluid
phases and therefore exhibits {\em two} gapless superfluid modes
displayed in Fig.~\ref{fig:phasediagram}c (with velocities $v_b$ and
$v_c$), observable via Bragg spectroscopy~\cite{bragg}.

Hence through the decoupling transition charge Luttinger parameters
``jump'' from $\bar{K}_{b,c}$ down to $K_{b,c}$. This is reflected in
the corresponding momentum distribution functions $n_{b,c}(k)$, that
for molecules [using Eq.~(\ref{eq:bs})] is given by $n_b(k) \propto \int
dx {\rm e}^{ikx}\langle b^\dagger(x) b(0)\rangle\sim k^{-\alpha}$ (for
$k\to 0$), with $\alpha_{\rm coupled} = 1-(2\bar{K}_b)^{-1}>
\alpha_{\rm decoupled} = 1-(2K_b)^{-1}$. This abrupt enhancement in
the low $k$ peak of $n_b(k)$ reflects the suppression of molecular
phase fluctuations in the coupled phase (due to locking to $\phi_c$)
and should be measurable via time-of-flight images~\cite{Paredes}.

Hence, as advertised in Fig.1, we predict a quantum phase transition
between the coupled, CSG and decoupled, DSG spin-gapped superfluids.
To determine the phase diagram outside of the deep spin-gap regime,
i.e., for values of the backscattering amplitude $\y$ other than
$\y\ll -1$, requires a detailed RG analysis.
As outlined above the RG computation involves progressively
integrating out (perturbatively in $\g$ and $\y$) degrees of freedom
at the UV cutoff scale $a^{-1}$. For simplicity we specialize here to
the case of equal velocities ($v_b \simeq v_c \simeq v_s$), leaving
the more technically challenging general case to a future
study~\cite{detailedPRB}. The result of this coarse-graining procedure
is summarized by the RG flow equations
\bea
\label{eq:RG1}
\frac{d \y}{d\ell} &=& 2\y(1- K_s) -\frac{1}{2}\g^2,
\\
\label{eq:RG2}
\frac{d \g}{d\ell}&=&(2-D_{\rm fr}-\frac{1}{2}K_s)\g-\frac{1}{2}\g \y,
\eea
\bea
\label{eq:RG3}
\frac{dK_s}{d\ell}&=&-\frac{1}{2} \y^2 -\frac{1}{4}\g^2,
\\
\label{eq:RG4}
\frac{d D_{\rm fr}}{d\ell}&=& -\frac{9}{8}\g^2.
\eea
One important subtlety~\cite{Giamarchi} is that abelian bosonization
does {\em not} explicitly capture the underlying SU(2) spin-rotation
symmetry embodied in our system's Hamiltonian. It can, however, be
restored (as we have done above) by a specific choice of the UV cutoff
$a$ and by imposing the relation $K_s = 1+ \y/2$ on the initial
conditions for the flows.  This constraint stems from the fact that
the value of $\y$ and the correction to $K_s$ arise from the same
fermion interaction. It is simple to verify that this SU(2) invariance
relation between $K_s$ and $\y$ is preserved by the RG flows
Eq.~(\ref{eq:RG1}) and~(\ref{eq:RG3}), i.e., that for $K_s = 1+ \y/2$
initially, $d(1+\y/2-K_s)/d\ell=\y(1+ \y/2-K_s)=0$.

Incorporating this relation significantly simplifies the RG flow
equations, giving:
\bea
\label{eq:RG1new}
\frac{d \y}{d\ell} &=& -\y^2 - \g^2/2,
\\
\frac{d\g}{d\ell} &=& (3/2 - D_{\rm fr} - 3\y/4 ) \g,
\label{eq:RG2new}
\eea
along with Eq.~(\ref{eq:RG4}) which is unchanged.  The phase diagram
Fig.~\ref{fig:phasediagram} is determined by the asymptotic (large
$\ell$, corresponding to low energies and long scales) flows of
$\g(\ell)$, $D_{\rm fr}(\ell)$ and $\y(\ell)$. A priori, one might
have expected four distinct phases corresponding to four different
combinations of relevent and irrelevant regimes of two couplings $\y$
and $\g$. However, as is clear from the structure of the flow
equations [particularly Eq.~(\ref{eq:RG1new})], it is impossible to
have a phase-coupled but spin-gapless state characterized by an
asymptotically nonzero $\g(\ell\rightarrow\infty)$ and a vanishing
$\y$. That is, back-scattering $\y$ (that leads to Cooper-pair singlet
formation) is always relevant when $\g$ is.  Physically this can be
understood by observing directly from $S$ that an arbitrary strength
FR coupling, at $T=0$ always induces pairing in an itinerant (i.e.,
ignoring Mott-insulating effects of a commensurate lattice) fermionic
atom system.

One simple limit of the flow equations is $D_{\rm fr}\gg 1$, in which
case $\g$ clearly flows to $0$ and the phase boundary separating the
DSG (for $\y<0$) and decoupled spin-gapless DNG (for $\y>0$) states
asymptotes to $\y=0$, respectively corresponding to attractive and
repulsive nonresonant atomic interactions.

Another observation is that, for large repulsive background
interactions $\y\gg 1$, the phase boundary separating the CSG and DNG
states asymptotes to $D_{\rm fr}=3/2$, as can be seen from the flow
equations. These show that a strong background repulsion strongly
suppresses the FR coupling $\g$ through Eq.~(\ref{eq:RG2new}), so that
it has little effect on the flow of $\y$, which then itself can flow
to $0$ along the $\g\approx 0$ sine-Gordon separatrix. This then
reduces the eigenvalue for $\g$ in Eq.~(\ref{eq:RG2new}) to
$3/2-D_{\rm fr}$ leading to the phase boundary $D_{\rm fr}=3/2$.
This is consistent with the observation that the FR coupling scaling
dimension $D = D_{\rm fr} + \frac{1}{2} K_s$ is $2$ at the transition
and $K_s=1$ in the DNG state, constrained by the spin-rotational
invariance.
For an intermediate and small positive values of $\y$, we have
numerically integrated the RG equations to determine the interpolation
of the vertical part ($D_{\rm fr}=3/2$) to the horizontal part
($\y=0$) of the phase boundary, as illustrated in
Fig.~\ref{fig:phasediagram}.

In contrast to repulsive background interactions that can be
irrelevant if the FR coupling is, attractive interaction ($\y<0$)
always grows under coarse-graining. Physically, this is a reflection
that Cooper-pairing always takes place at $T=0$, even in the absence
of the FR coupling that can only enhance spin-singlet formation.
Hence, at sufficiently long scales, the RG flows leave their
perturbative regime of validity confined to $|\y|\lesssim 1$ and
$|\g|\lesssim 1$. Outside of this range requires a nonperturbative
analysis.  As discussed earlier, the deep spin-gap state $\y\ll -1$
can nevertheless be analyzed by simply setting $\theta_s=0$ in the
action and recomputing the flows perturbatively in $\g$. As
illustrated in Fig.~\ref{fig:phasediagram}, this leads to a phase
boundary between CSG and DSG that is asymptotically vertical and given
by $D_{\rm fr}=2$.

To calculate this phase boundary for more moderate (but still
negative) values of $\y$ requires a \lq\lq matching\rq\rq\ 
calculation.  To see this we note that for small negative $\y$,
perturbative RG flows (describing spin fluctuations on shorter length
scales inside the spin-gapped state), remain valid up to a crossover
spin-gap length scale, $\xi_{bs}\approx a v_s^{1/2}
|\y|^{-1/(2-2K_s)}$ (coinciding with the width of the soliton in the
sine-Gordon model).  Hence we integrate the perturbative RG flows out
to length $\xi_{bs}= a v_s^{1/2} e^{\ell_*}$, defined by
$\y(\ell^*)\approx 1$, beyond which spin fluctuations freeze out,
$\theta_s\approx 0$. On longer scales we match onto the strongly
coupled spin-gapped state, characterized by setting $\theta_s=0$
inside the action, Eq.~(\ref{eq:wholeaction}) but with the
fluctuation-renormalized parameters $\g(\ell^*)$ and $D_{\rm
  fr}(\ell^*)$ determined by the RG flow.  Setting $D_{\rm
  fr}(\ell^*)=2$ then determines the fluctuation-renormalized CSG-DSG
phase boundary illustrated in Fig.~\ref{fig:phasediagram}.  The shift
of the phase boundary toward a larger critical $D_{\rm fr}$ can be
seen from the decrease of $D_{\rm fr}(\ell)$ under coarse-graining,
which in turn corresponds to enhancement of superfluidity by the FR
coupling.

A nontrivial challenge is the experimental realization of a LL with
parameters tunable across the decoupling transition.  The necessary
large value of $D_{\rm fr}>3/2$, requires small Luttinger parameters,
$K_\alpha$, realized by strongly interacting systems, for which a
relation to microscopic parameters is difficult to establish. One
exception is the extended quarter-filled 1d Hubbard model, where in
the limit of large on-site repulsion, $K_c = [2+ 4/\pi \sin^{-1} v
]^{-1}$, with $v=V/(2|t|)$ a ratio (limited to $<1$ by phase
separation) of the nearest neighbor hopping $t$ and repulsion $V$
energies, respectively~\cite{Schulz90}.  Hence, $K_c$ as low as a
$1/4$ can be reached, assuring a transition (that for strong repulsion
takes place at $D_{\rm fr}=3/2$) into the decoupled superfluid state
at $v = 1/\sqrt{2}$ for weakly interacting molecules. In the more
favorable molecular Tonks regime with $K_b\rightarrow 1$, we predict
the quantum transition at $v\approx0.38$.  Thus we suggest that a
Feshbach-resonantly interacing atomic gas confined in a highly
anisotropic (1d) trap and subjected to a periodic optical
potential~\cite{Paredes} is a promising candidate for an
experimental realization of the phase diagram and the decoupling
transition discussed here.

We gratefully acknowledge discussions with Victor Gurarie and Matthew
Fisher, as well as support from NSF DMR-0321848 and the Packard
Foundation.
\vspace{-0.2cm}


\begin{thebibliography}{10}
\bibitem{comment} This is a reasonable assumption even in 1d in most
  currently realized experimental systems\cite{expts} characterized by
  a strong FR coupling (i.e., with width much larger than the Fermi
  energy), so that closed and open channels are strongly hybridized
  precluding an unabiguous separation between molecules and Cooper
  pairs. The latter is, however, possible for narrow FR\cite{Hulet}.
\end{thebibliography}
\end{document}